\documentclass[aps,prb,twocolumn,superscriptaddress]{revtex4-1}
\usepackage{graphicx}

\usepackage{amsmath}
\usepackage{amssymb}
\usepackage{graphicx}
\usepackage{afterpage}
\usepackage{amsfonts}
\usepackage{amssymb}
\usepackage{graphicx}
\usepackage{braket}

\newcommand{\be}{\begin{equation}}
\newcommand{\ee}{\end{equation}}
\newcommand{\bea}{\begin{eqnarray}}
\newcommand{\eea}{\end{eqnarray}}

\newcommand{\Ea}{\ensuremath{{\cal E}_1}}
\newcommand{\Eb}{\ensuremath{{\cal E}_2}}
\newcommand{\Ec}{\ensuremath{{\cal E}_3}}
\newcommand{\Ei}{\ensuremath{{\cal E}_i}}
\newcommand{\Ef}{\ensuremath{{\cal E}_{\boldsymbol{f}}}}
\newcommand{\En}{\ensuremath{{\cal E}_{\boldsymbol{f'}}}}
\newcommand{\Es}{\ensuremath{{\cal E}_{\boldsymbol{s}}}}
\newcommand{\Eg}{\ensuremath{{\cal E}_{\boldsymbol{n}}}}
\newcommand{\Ed}{\ensuremath{{\cal E}_{QD}}}
\newcommand{\Eo}{\ensuremath{{\cal E}_{1,\,2,\,3}}}
\newcommand{\Er}{\ensuremath{{\cal E}_{r}}}

\renewcommand{\dag}{^{\dagger}}

\begin{document}


\title{Multi-Wave Coherent Control of a Solid State Single Emitter}



\author{F. Fras}
\email{francois.fras@ipcms.unistra.fr} \affiliation{Univ. Grenoble
Alpes, F-38000 Grenoble, France} \affiliation{CNRS, Institut
N\'{e}el, F-38000 Grenoble, France} \affiliation{\,IPCMS UMR 7504,
CNRS and Universit\'{e} de Strasbourg, Strasbourg, France}

\author{Q. Mermillod}
\author{G. Nogues}
\author{C. Hoarau}
\affiliation{Univ. Grenoble Alpes, F-38000 Grenoble, France}
\affiliation{CNRS, Institut N\'{e}el, F-38000 Grenoble, France}

\author{C. Schneider}
\author{M. Kamp}
\affiliation{Technische Physik and Wilhelm Conrad R\"{o}ntgen
Research Center for Complex Material Systems, Universit\"{a}t
W\"{u}rzburg, Germany}

\author{S. H\"{o}fling}
\affiliation{Technische Physik and Wilhelm Conrad R\"{o}ntgen
Research Center for Complex Material Systems, Universit\"{a}t
W\"{u}rzburg, Germany} \affiliation{SUPA, School of Physics and
Astronomy, University of St Andrews, St Andrews, KY16 9SS, United
Kingdom}

\author{W. Langbein}
\affiliation{Cardiff University School of Physics and Astronomy, The
Parade, Cardiff CF24 3AA, United Kingdom}

\author{J. Kasprzak}
\email[]{jacek.kasprzak@neel.cnrs.fr} \affiliation{Univ. Grenoble
Alpes, F-38000 Grenoble, France} \affiliation{CNRS, Institut
N\'{e}el, F-38000 Grenoble, France}


\maketitle

\date{\today}

{\bf Coherent control of individual two-level systems (TLSs) is at
the basis of any implementation of quantum information. An
impressive level of control is now achieved using
nuclear\,\cite{PlaNature13, MuhonenNatNano14},
vacancies\,\cite{WidmannNatMat15, YalePNAS13} and charge
spins\,\cite{CarterNatPhoton13, HansomNatPhys14}. Manipulation of
bright exciton transitions in semiconductor quantum dots (QDs) is
less advanced, principally due to the sub-nanosecond
dephasing\,\cite{BorriPRL01}. Conversely, owing to their robust
coupling to light, one can apply tools of nonlinear
spectroscopy\,\cite{LangbeinOL06} to achieve all-optical command.
Here, we report on the coherent manipulation of an exciton via
multi-wave mixing. Specifically, we employ three resonant pulses
driving a single InAs QD. The first two induce a four-wave mixing
(FWM) transient, which is projected onto a six-wave mixing (SWM)
depending on the delay and area of the third pulse, in agreement
with analytical predictions. Such a switch enables to demonstrate
the generation of SWM on a single emitter and to engineer the
spectro-temporal shape of the coherent response originating from a
TLS. These results pave the way toward multi-pulse manipulations of
solid state qubits via implementing the NMR-like control
schemes\,\cite{VandersypenRMP05} in the optical domain.}

An appealing strategy to realize optically controlled quantum
networks in solid state, is to coherently couple distant TLSs - like
bright exciton transitions in QDs - via photons confined in
microcavities\,\cite{ReithmaierN04, AlbertNatComm13} or propagating
in waveguides\,\cite{LuxmoorePRL13, ColesOptEx14,
ReichertPhysRevB14, ArcariPRL14}. The efficient retrieval and
manipulation of coherent responses from single excitons is mandatory
for assessing the properties of such solid state qubits and
demonstrating the coherence transfer between them. In this context,
substantial progress has been made by the introduction of the
heterodyne spectral interferometry technique\,\cite{LangbeinOL06} to
measure four-wave mixing (FWM) from individual excitons in various
nanostructures\,\cite{PattonPRL05, KasprzakNMa10, KasprzakNPho11,
AlbertNatComm13, KasprzakNJP13}. Yet, measuring on single
strongly-confined excitons in InAs QDs remains challenging and has
not been previously achieved. This is due to their small dipole
moment $\mu$ and the resulting required high resonant amplitudes of
the three fields; $\Ea,\,\Eb,\,\Ec$, employed to drive the measured
FWM polarization, being proportional to $\mu^4\Ea^\ast\Eb\Ec$ and
higher order terms with the same phase dependence on the fields.

Here, the retrieval and manipulation of wave-mixing signals from
single InAs QDs is accomplished by embedding them in a low-Q planar
semiconductor microcavity\,\cite{MeierOptEx14}, as detailed in
Supplementary Fig.\,S2. Such a semiconductor hetero-structure
provides an intra-cavity field enhancement of $\sqrt{Q}\simeq13$ at
the QDs' position, whilst offering spectral matching of the photonic
mode with femto-second laser pulses. This cavity-enhanced driving
allows reducing the resonant excitation intensity, required to
induce the FWM, by a factor around $Q^{3/2}\simeq2200$. In
consequence, the retrieval efficiency of wave-mixing signals from
single excitons is improved by two orders of magnitude with respect
to previous experiments on bare QDs\,\cite{LangbeinOL06,
KasprzakNJP13}. In this work, we employ such wave-mixing responses
to realize a novel scheme for ultrafast coherent control of
optically active TLSs. We demonstrate gating of their coherent
emission, by converting the FWM polarization into the SWM one. We
show engineering of the FWM spectral response, also acting on the
global spectral lineshape of a TLS. Details regarding the current
experimental configuration are given in Supplementary Fig.\,S1.

\begin{figure}[!ht]
\includegraphics[width=1.02\columnwidth]{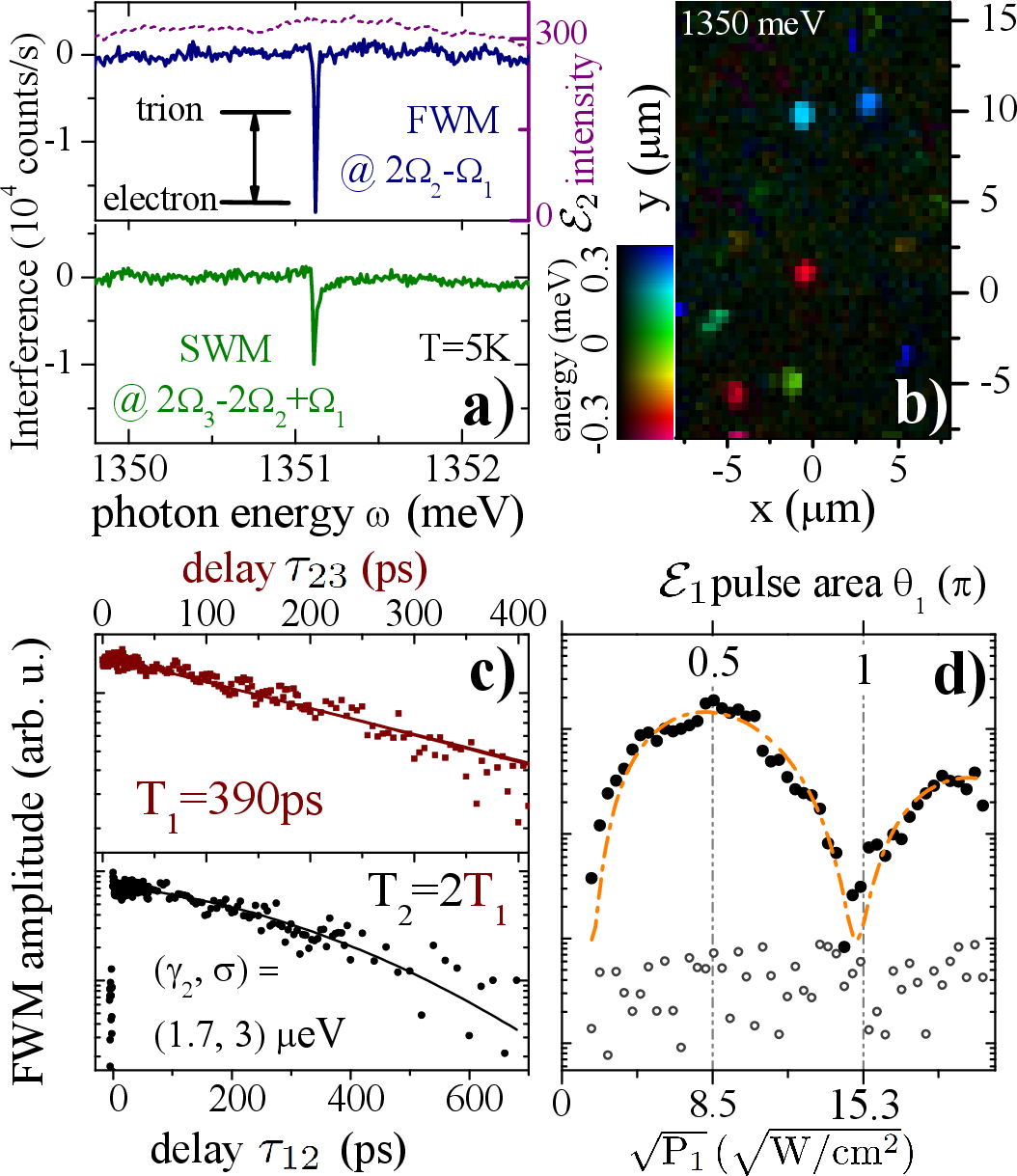}
\caption{\textbf{Wave mixing spectroscopy of individual negative
trions in InAs QDs embedded in a low-Q microcavity.}
a)\,Interferograms of four-wave mixing (FWM) of a single trion for
$2\Omega_2-\Omega_1$, $\tau_{12}=0.5\,$ps and pulse areas
$(\theta_1,\,\theta_2)=(\pi/2,\,\pi)$, and six-wave mixing (SWM) at
$2\Omega_3-2\Omega_2+\Omega_1$ for $\tau_{12}=0.5\,$ps,
$\tau_{23}=1\,$ps,
$(\theta_1,\,\theta_2,\,\theta_3)=(\pi/2,\,\pi,\,\pi)$. Integration
time 20\,s. Heterodyned spectral shape of $\Eb$ is shown with a
purple, dashed line. b)\,Amplitude of the FWM hyperspectral imaging
at $\tau_{12}=1\,{\rm ps}$. Localized peaks correspond to the FWM of
single QDs: the amplitude is displayed as brightness, the energy is
indicated as the hue of the colour range (indicated by the bar) from
-0.3 (red) to +0.3\,meV (blue) with respect to the chosen center of
$\hbar\omega=1350\,$meV. c)\,Population dynamics (brown squares) of
the trion probed with FWM, $\tau_{12}=0.5\,{\rm ps}$: exponential
decay (brown line) yields the population lifetime of ${\rm
T}_1=(390\,\pm\,10)\,$ps. Coherence dynamics (black circles) of the
same trion probed with FWM, consistent with radiatively limited
dephasing T$_2$=2T$_1$ and a residual inhomogeneous broadening of
$\sigma=(3\,\pm\,1)\,\mu$eV. d)\,FWM power as a function of
$\sqrt{{\rm P}_1}=|\Ea|$ and of $\Ea$ pulse area $\theta_1$ showing
Rabi oscillations, $\theta_2=2/3\pi$, ${\rm P}_2=|\Eb|^2=157\,{\rm
W}/{\rm cm^{2}}$, $\tau_{12}=0.5\,{\rm ps}$. Pulse areas of
$(\theta_1, \theta_2)=(\pi/2,\,\pi)$, corresponding to $({\rm
P}_1,\,{\rm P}_2)=(72,\,235)\,$W/cm$^{2}$, are determined by the
first maximum of the FWM versus $\theta_1$ ($\theta_2$, not shown).
Predicted sin$^2(\theta_1)\exp(-\xi\theta_1)$ is depicted by the
orange dashed-dotted line, where $\xi$ is the damping constant.
\label{fig:WM_spec}}
\end{figure}

For the purpose of the experiment described in this Letter, we
require an optically active TLS in a solid. In InAs QDs, this is the
case of negatively charged excitons (trions)\,\cite{BennyPRB12},
where the level structure can be trimmed down to two levels
(disregarding spin degeneracy): the ground and excited states
corresponding to the presence of an electron and a trion in the QD,
respectively, as depicted in Fig.\,\ref{fig:WM_spec}\,a. We have
therefore used a sample which is intentionally
n-doped\,\cite{MeierOptEx14} with a Silicon $\delta$-layer located
in the cavity spacer. To illustrate the coherent response of trions,
we present in Fig.\,\ref{fig:WM_spec}\,a the spectral interferogram
detected at the FWM frequency $2\Omega_2-\Omega_1$, as detailed in
the Supplementary Material. Based on series of interferograms, we
construct the FWM hyperspectral imaging\,\cite{KasprzakNPho11}, as
exemplified in Fig.\,\ref{fig:WM_spec}\,b. Therein, the brightest,
localized peaks correspond primarily to the FWM generated by trions.
Such imaging is employed to determine spatial and spectral position
of QDs. Also it enables a detailed statistical analysis on excitonic
complexes, as exemplified for biexcitons in Supplementary Fig.\,S3.

We first retrieve the required parameters to infer and model the
optical response of single emitters in a solid. The three-beam FWM
spectroscopy offers direct access to coherence and population
dynamics in the TLS. To probe the population dynamics, we detect the
time-integrated FWM at the $\Omega_3+\Omega_2-\Omega_1$ frequency,
while varying the delay $\tau_{23}$ between $\Eb$ and $\Ec$ and
fixing $\tau_{12}=0.5\,$ps. The resulting evolution of the FWM
amplitude is shown in Fig.\,\ref{fig:WM_spec}\,c and fitted by an
exponential decay, yielding\,\cite{LangbeinPRB04a,ProuxPRL15} the
lifetime of T$_1=(390\,\pm\,10)\,$ps. Conversely, detecting the FWM
at the frequency $2\Omega_2-\Omega_1$, reflects the coherence in the
TLS. Its evolution is governed by the delay $\tau_{12}$ between
$\Ea$ and $\Eb$, as displayed in Fig.\,\ref{fig:WM_spec}\,c. The
lack of FWM at negative delays and the absence of fine-structure
beating\,\cite{KasprzakNJP13} confirm the trionic nature of the
investigated transition. Our data are well described by the product
of an exponential and a Gaussian decay\,\cite{KasprzakNJP13}. From
the former, we infer the dephasing time T$_2$ and the related
homogenous broadening $\gamma_2=2\hbar/{\rm T}_2$ (FWHM). The latter
yields the inhomogeneous broadening $\sigma$ due to the residual
spectral wandering, occurring during the integration time. Note that
even on the single transition level such wandering creates a photon
echo, manifested here by a Gaussian decay of the
coherence\,\cite{PattonPRB06, KasprzakNJP13}. The coherence dynamics
can be fitted using T$_2$=2T$_1$=780\,ps and
$\sigma=(3\,\pm\,1)\,\mu$eV. We thus find that the coherence of
trions in these QDs approaches the radiative limit. This is
supported by observation of their first order reflectance, as shown
in Supplementary Fig.\,S5. More examples of such transitions close
to the radiative limit are given in Supplementary Fig.\,S4.

\begin{figure}[!ht]
\includegraphics*[width=1.02\columnwidth]{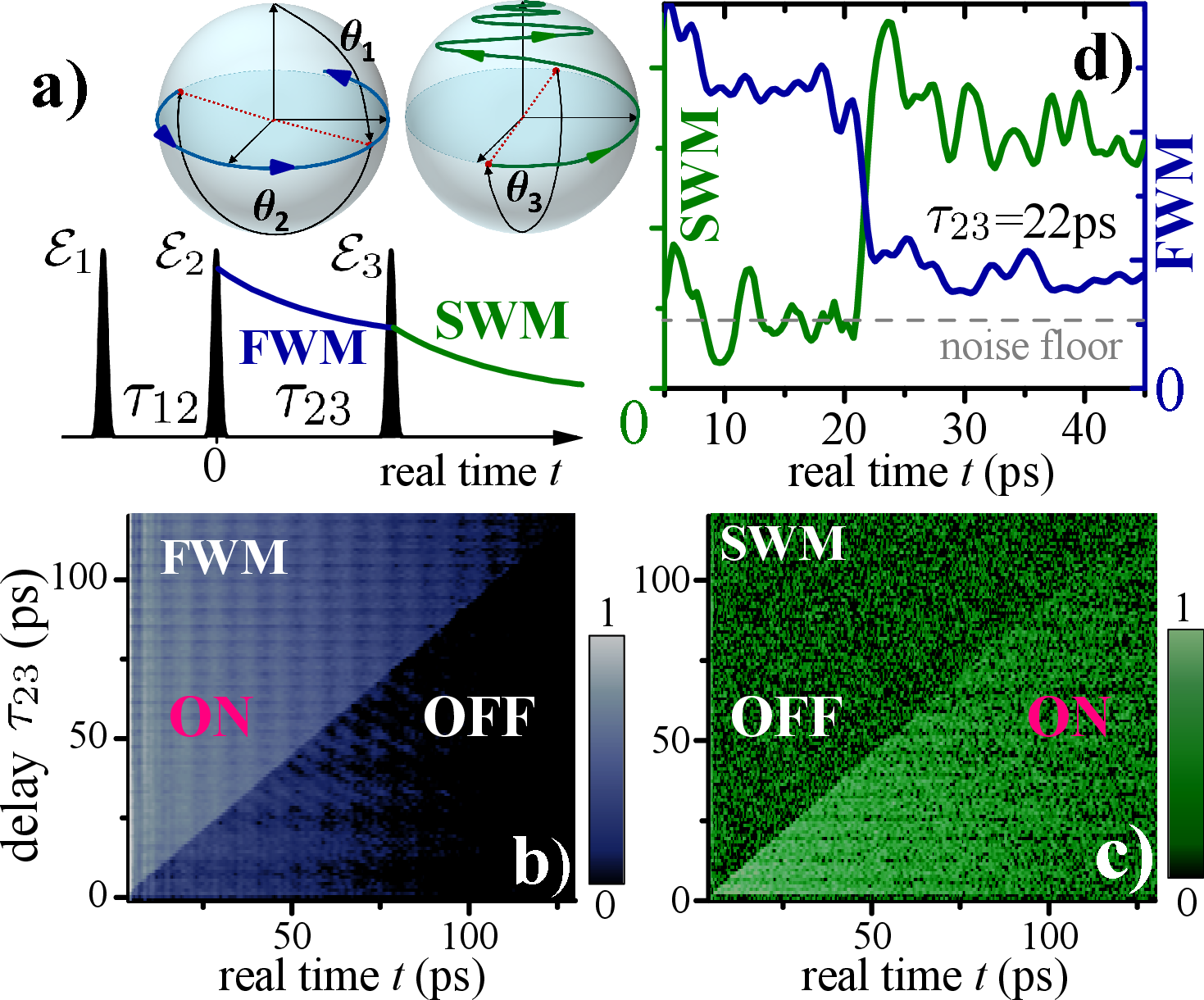}
\caption{\textbf{FWM/SWM switching.} a)\,Pulse sequence used in the
experiment, highlighting rationale of the FWM/SWM switching. The
evolution of the trion's Bloch vector (red) during FWM, switching
and SWM is also depicted (ground state on the top of the sphere).
The FWM transient (blue line) is created by $\Ea$ and $\Eb$ fields
arriving at $t=-\tau_{12}$ and $t=0$, respectively and it freely
evolves during $\tau_{23}$. $\Ec$, arriving at $t=\tau_{23}$,
converts the FWM into the SWM (green line). The FWM/SWM conversion
efficiency reaches unity for $\theta_3=\pi$. FWM (b) and SWM (c)
amplitudes as a function of real time $t$ and delay $\tau_{23}$ for
$\theta_3=0.8\pi$. FWM (SWM) signals are present above (below) the
diagonal giving the arrival of $\Ec$. Amplitude on a linear colour
scale, as given. Decreased amplitudes for long times is due to the
limited resolution of the spectrometer. d)\,Demonstration of the FWM
suppression (blue) and the SWM build up (green trace) at the arrival
of $\Ec$ for $t=\tau_{23}=22\,$ps. The noise floor is given by a
gray dashed line. Note the coexistence of both signals for
$t>22\,$ps and corresponding residual signal in
Fig.\,\ref{fig:FWM_control}\,c, owing to
$\theta_3\neq\pi$.\label{fig:FWM_SWM}}
\end{figure}

The coherent control experiment described below requires a
calibration of the pulse areas $\theta_i=\int dt \, \mu
|\Ei(t)|/\hbar $, which are proportional to the square root of the
pulse intensities ${\rm P}_i$. To illustrate this calibration, we
present in Fig.\,\ref{fig:WM_spec}\,d the FWM power as a function of
$\sqrt{{\rm P}_1}$ and $\theta_1$. As
expected\,\cite{BonadeoScience98,StievaterPRL01,PattonPRL05}, the
FWM undergoes Rabi oscillations with increasing $\theta_1$. In
addition, we observe a $\theta_1$-dependent damping, which is
attributed to dissipative coupling with acoustic
phonons\,\cite{RamseyPRL10}. In the three-beam heterodyne spectral
interferometry technique, the additional degree of freedom provided
by the time delay $\tau_{23}$ allows to transform the FWM into SWM
at the defined time after creating the FWM. Such SWM signals and
their interferences with coexisting FWM have been studied in atomic
physics\,\cite{ZhangOL07,ZhangPRL09}. Higher order wave-mixing has
been also investigated in the condensed matter physics to explore
many particle correlations in quantum wells\,\cite{TurnerN10,
AxtPRB01, VossPRB02} and QD ensembles\,\cite{MoodyPRB13}, recently
inferring their non-Markovian dynamics\,\cite{TaharaPRL14}. In the
present study, we demonstrate for the first time generation of SWM
on single emitters. Here, SWM detection serves as a tool to
implement coherent control upon the FWM transient of a TLS. We use
$\Ec$ pulse to project the FWM into the SWM signal oscillating at
the $2\Omega_3-2\Omega_2+\Omega_1$ frequency. The spectral
interferogram measured at this frequency is shown in
Fig.\,\ref{fig:WM_spec}\,a, for the applied pulse sequence given in
Fig.\,\ref{fig:FWM_SWM}\,a. The FWM/SWM swapping is controlled by
adjusting the pulse area $\theta_3$ and $\tau_{23}$, as derived in
the analytical model presented in the Supplementary Material: FWM
field ${\cal E}_{\boldsymbol{f}}$ and SWM field ${\cal
E}_{\boldsymbol{s}}$ read:
\begin{align}
\label{equ1}
\lvert {\cal E}_{\boldsymbol{f}}(t) \rvert  & \propto \sin{\theta_1}\sin^2{\frac{\theta_2}{2}} \left(   \mathrm{\Theta} (t)\vphantom{\frac{\theta_1}{2}} \right.\left.  -\sin^2{\frac{\theta_3}{2}} \mathrm{\Theta} (t-\tau_{23}) \right) \times \nonumber\\
  &e^{-t \gamma_2 },\\
\label{equ2} \lvert {\cal E}_{\boldsymbol{s}}(t)  \rvert & \propto
\sin{\theta_1} \sin^2{\frac{\theta_2}{2}} \sin^2{\frac{\theta_3}{2}}
\mathrm{\Theta} (t-\tau_{23})e^{-t \gamma_2 },
\end{align}

where $\Theta$ is the Heaviside function, and $t=0$ is defined by
the arrival of $\Eb$.

Conversion of FWM into SWM and their coexistence is experimentally
demonstrated in Fig.\,\ref{fig:FWM_SWM}. Therein, we present the FWM
and SWM field amplitude transients for increasing $\tau_{23}$ at
$\theta_3=0.8\pi$. In such maps, the arrival of $\Ec$ at
$t=\tau_{23}$, generating the conversion, defines the diagonal. From
Eqs.\,(\ref{equ1}) and (\ref{equ2}), we expect that the FWM is
present in the upper-side of the diagonal $(t<\tau_{23})$, and has
been converted to SWM in the lower-side $(t>\tau_{23})$, as indeed
measured. This temporal gating of both signals is the key to
manipulate the spectral distribution of the FWM from a TLS.

\begin{figure}[!ht]
\includegraphics[width=1.02\columnwidth]{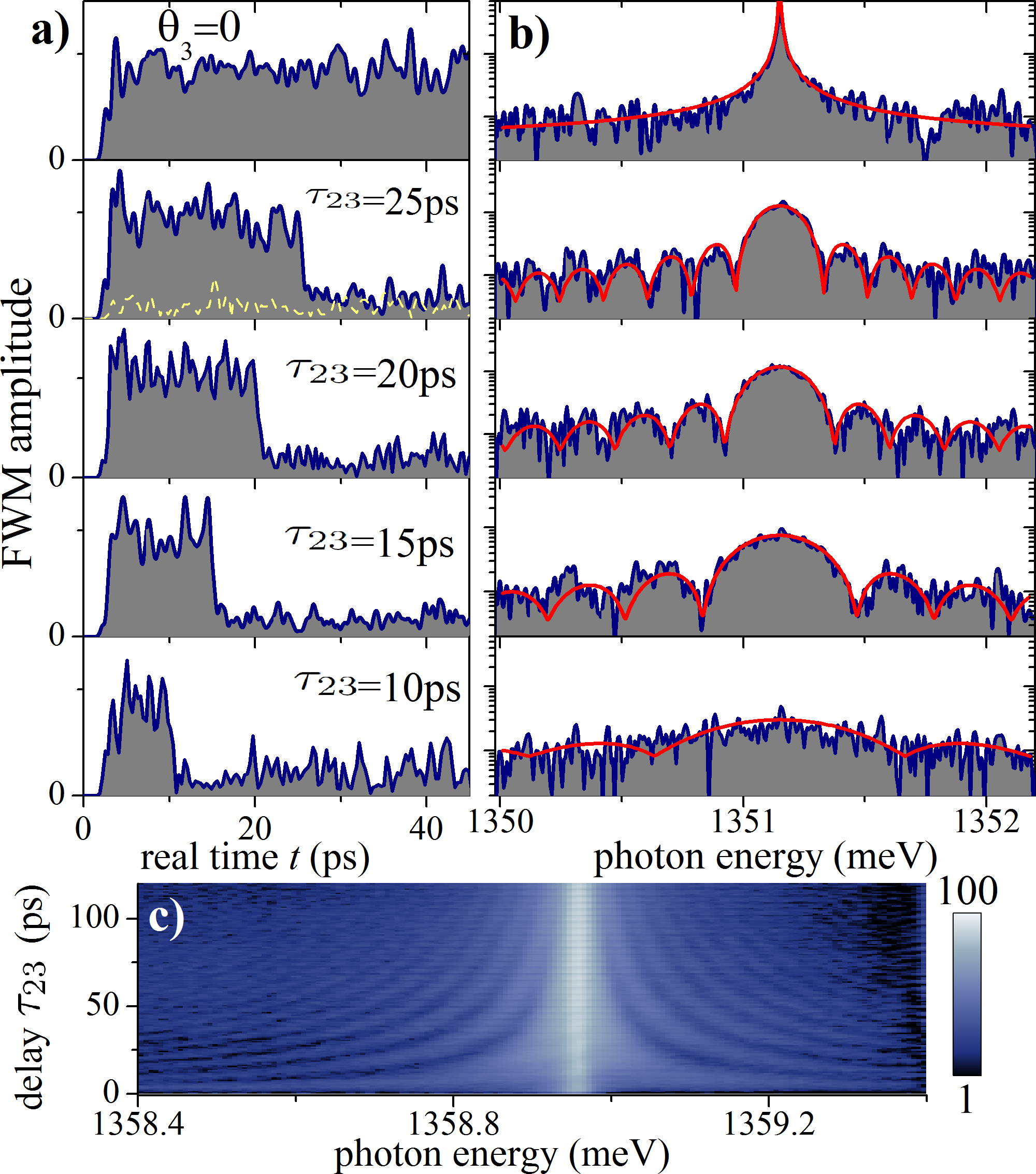} \caption{\textbf{Manipulation of the coherent response
of a single emitter via wave-mixing switching.} a)\,FWM transients
generated for $\tau_{12}=0.2\,{\rm ps}$, $\theta_3=\pi$ and
$\tau_{23}$ as labeled. A complete suppression of the FWM is
observed after the arrival of $\Ec$ ($t>\tau_{23}$); the noise level
is shown by the dashed yellow line. The free evolution of the FWM
for $\theta_3=0$ is shown in the topmost panel. b)\,FWM spectral
amplitudes of the data in a), showing a substantial spectral
broadening of the FWM due to the step-like suppression of the signal
in time domain. Analytical predictions using Eq.\,(\ref{TF2}) are
shown as red lines. c)\,FWM spectral amplitudes of the transients
given in Fig.\,\ref{fig:FWM_SWM}\,b, showing the evolution of the
FWM lineshape with increasing $\tau_{23}$ for $\theta_3=0.8\pi$.
Logarithmic colour scale, as shown by the vertical
bar.\label{fig:FWM_control}}
\end{figure}

We note that the conversion efficiency reaches unity for the
$\theta_3=\pi$ yielding, at the arrival of $\Ec$, a complete
suppression of the FWM. In Fig.\,\ref{fig:FWM_control}, we employ
specific pulse areas $(\theta_1,\theta_2)=(\frac{\pi}{2},\pi)$ in
order to drive, between $\Eb$ and $\Ec$, the maximum polarization to
the FWM, consistent with Eq.\,(\ref{equ1}). In
Fig.\,\ref{fig:FWM_control}\,a we present FWM transients measured at
$\theta_3=\pi$ for various $\tau_{23}$. Their spectral amplitudes
(also see Supplementary Fig.\,S6) are given in
Fig.\,\ref{fig:FWM_control}\,b. The spectra show a substantial
broadening and ringing due to the step-like suppression in time,
leading to a sine cardinal like shape. FWM at $\theta_3=0$ (topmost)
is given for comparison. The effect, also clearly visible in
Fig.\,\ref{fig:FWM_control}\,c, is observable from
$\tau_{23}=110\,$ps to $\tau_{23}=10\,$ps where the FWHM of the main
peak increases from 38\,$\mu$eV (given by the spectral resolution)
up to 700$\,\mu$eV. This represents an imposed broadening by two
orders of magnitude with respect to $\gamma_2$. With further
decrease of $\tau_{23}$ such spectral broadening can reach 10\,meV
range, as is only bounded by the duration of $\Ec$. Due to the
limited amount of the emitted FWM for $\tau_{23}\simeq0$, its
observation requires a large signal-to-noise ratio. Fourier
transforming Eq.\,(\ref{equ1}) with respect to the real time
\textit{t}, yields the spectrally resolved FWM amplitude:
\begin{align}
\label{TF2}
&\lvert {\cal E}_{\boldsymbol{f}}(\omega) \rvert \propto \sin{\theta_1}\sin^2{\frac{\theta_2}{2}}e^{-\tau_{12}\gamma_2} \times \nonumber\\
&\left( \frac{ 1 +\sin^4{\frac{\theta_3}{2}}e^{-2\tau_{23}\gamma_2}
-2\sin^2{\frac{\theta_3}{2}}e^{-\tau_{23}\gamma_2}\cos{\Delta \omega
\tau_{23}}  }{2\pi(\gamma_2^2+\Delta \omega^2)} \right)^{1/2}
\end{align}

where $\Delta\omega=\omega-\omega_{eg}.$ With the particular pulse
areas
$(\theta_1,\,\theta_2,\,\theta_3)=(\frac{\pi}{2},\,\pi,\,\pi)$, the
spectrum takes the simple form  $e^{-\gamma_2 (\tau_{12}+\tau_{23})}
\left(
\frac{\cosh({\gamma_2\tau_{23}})-\cos{(\Delta\omega\tau_{23})}}{\pi(\gamma_2^2+\Delta\omega^2)}
\right)^{1/2}$, which reproduces the observed features
quantitatively, as shown by red traces in
Fig.\,\ref{fig:FWM_control}\,b. The spectral broadening of the
central peak scales as $\frac{2\pi\hbar}{\tau_{23}}$, characteristic
of the sine cardinal lineshape.

\begin{figure}[!ht]
\includegraphics[width=1.02\columnwidth]{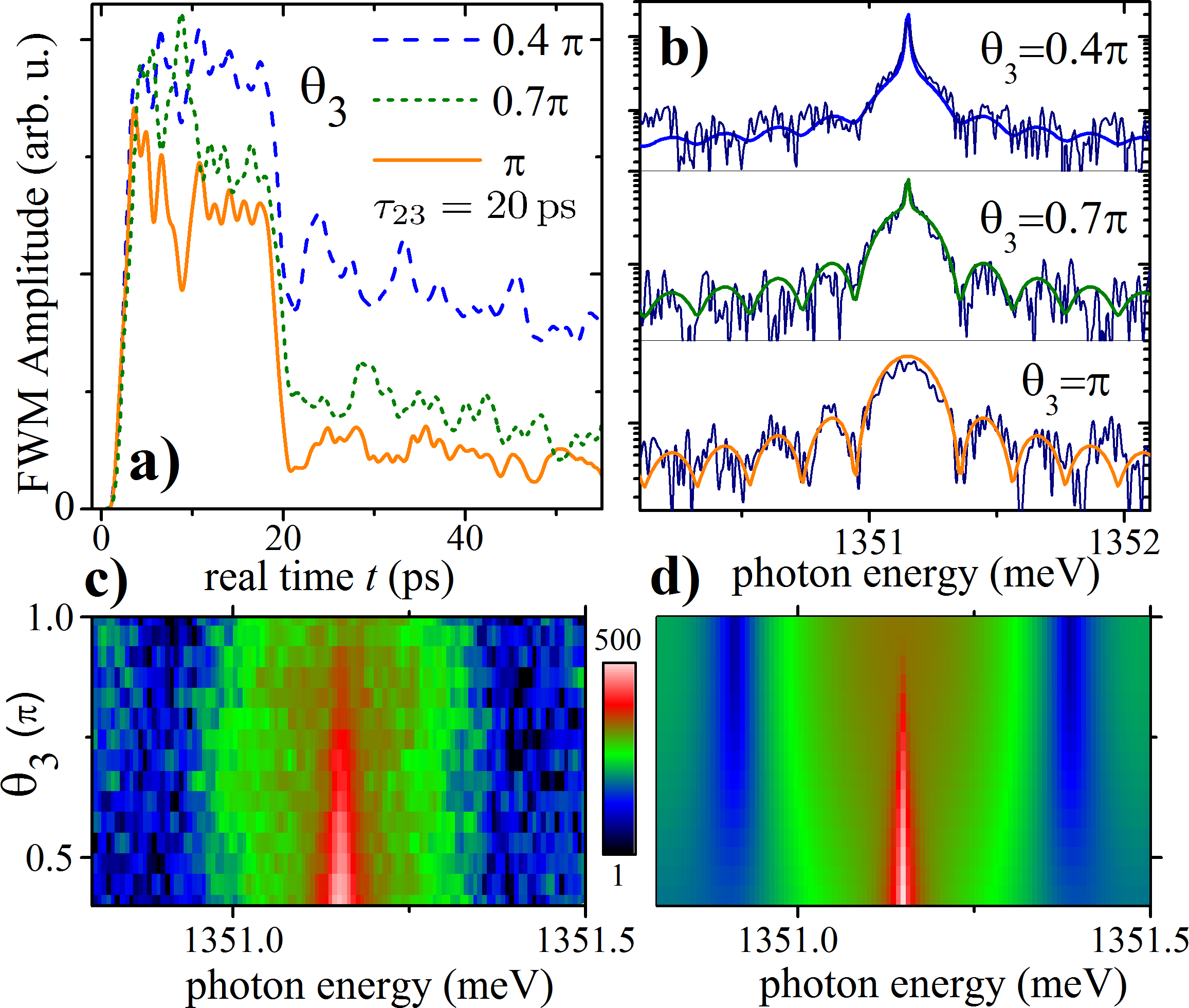}
\caption{\textbf{Manipulation of the FWM with the area of the
control pulse $\Ec$.} a)\,Measured evolution of the FWM transient
for $\theta_3$ of $0.4\pi$ (blue dashed), $0.7\pi$ (green dotted)
and $\pi$ (orange solid); for $\tau_{12}=0.2\,{\rm ps}$ and
$\tau_{23}=20\,{\rm ps}$. An increasing suppression of the FWM for
$t>\tau_{23}$ with increasing $\theta_3$ is observed. b)\,FWM
spectral amplitudes of the transients shown in a), with the
corresponding theoretical prediction according to Eq.\,(3). c)\,Map
of the FWM spectral amplitude as a function of $\theta_3$
demonstrating a gradual suppression of the spectrally narrow
component of the FWM. Delays as in a). d)\,Theoretical simulation of
c). Logarithmic color scale, as shown with the colour bar.
\label{fig:theta_control}}
\end{figure}

For intermediate values of $\theta_3 $, the FWM is only partially
converted to SWM. Therefore both non-linearities coexist, as can be
noted in Fig.\,\ref{fig:FWM_SWM}\,d. This is investigated in
Fig.\,\ref{fig:theta_control}, where the FWM transients (a) and
spectra (b) for increasing $\theta_3$ and fixed $\tau_{23}=20\,$ps
are presented. Owing to the increasing FWM suppression for
$t>20\,$ps, the amplitude of the spectrally narrow component is
reduced (see Supplementary Fig.\,S7). In parallel, the broad
pedestal develops. Hence, non-natural lineshapes can be designed by
tuning $\theta_3 $ and $\tau_{23}$. Eq.\,(\ref{TF2}), derived in the
Supplementary Material, reproduces the experimental data shown in
Fig.\,\ref{fig:theta_control}\,b and c, without any free parameters,
apart from the absolute common scaling.

The ability to manipulate the spectral width of a coherent response
from the TLS provides a novel degree of freedom in solid state
quantum optics. We emphasize that the phase shift induced by the
FWM/SWM conversion modifies the coherent oscillation of the dipole:
the global lineshape of the TLS polarization is altered, as
highlighted in Supplementary Fig.\,S8. The latter is strikingly
manifested when the $\Ec$ is heterodyned at $2\Omega_2-\Omega_1$,
leading to a stationary phase shift between the two non-linearities
generated by $\Eb$ and $\Ec$, respectively. Therefore by adjusting
this phase shift with $\tau_{23}$ (see Eqs.\,(11) and (12) in the
Supplementary Material), we can perform controlled phase rotations
of the TLS dipole. After arrival of $\Ec$, the phase of the TLS
emission could be inverted or entirely frozen. Such on-demand
blockade of emission represents a fundamental step in optical
control of single TLS systems. As an example of application, let us
note that tuning the spectral shape of a TLS could be employed to
optimize injection and storing of single photons in optical
resonators. Furthermore, maximizing the spectral overlap within a
pair of distant TLSs promotes their coherent coupling via
propagating photons\,\cite{MinkovPRB13}, which is a prerequisite to
realize optically controlled quantum networks in a solid. Multi-wave
mixing could be also used to selectively address the coherence
dynamics and transitions in higher manifolds of the Jaynes-Cummings
ladder of nanophotonic devices operating in a strong-coupling
regime\,\cite{KasprzakNMa10, AlbertNatComm13}.

By employing a low-Q semiconductor microcavity, we improved the
retrieval sensitivity of coherent responses from individual
emitters, enabling wave mixing spectroscopy of individual excitons
in strongly-confined InAs QDs. Using three beam configuration, we
inferred a new scheme for controlling coherent evolution of a TLS
via converting FWM to SWM. We have demonstrated that, via temporal
gating of the FWM, we can control its lineshape by varying the
amplitude and the arrival time of the gate pulse. The demonstration
of SWM on individual emitters paves the way towards investigations
of their non-Markovian dynamics\,\cite{TaharaPRL14} and to study
higher order correlations\,\cite{TurnerN10} involving more than two
excitons within a QD. Extending our proof-of-principle protocol
toward control via multi-wave mixing is conceptually
straightforward: conversion between wave mixing processes is
determined by areas and respective delays of driving pulses. It is
also technically feasible via multiplexed digital heterodyning. An
alluring perspective is to perform multi-wave mixing on radiatively
coupled pairs of distant excitons\,\cite{MinkovPRB13}, achieving
non-local quantum control in a solid.

\textit{We acknowledge support by the European Research Council
Starting Grant ``PICSEN" contract no.\,306387.}

\newpage

\widetext

\begin{center}
{\bf \large SUPPLEMENTARY MATERIAL\\ Multi-Wave Coherent Control of
a Solid State Single Emitter\\} F. Fras, Q. Mermillod, G. Nogues, C.
Hoarau, C. Schneider,\\ M. Kamp, S. H\"{o}fling, W. Langbein, and J.
Kasprzak
\end{center}

\setcounter{page}{1} \setcounter{figure}{0}

\renewcommand{\figurename}{Supplementary Figure}
\renewcommand{\thefigure}{S\arabic{figure}}

\section{Theory: Derivation of the FWM and SWM responses}

In this part we derive the four-wave mixing (FWM) and six-wave
mixing (SWM) responses, as experimentally studied in the main part
of the manuscript. We consider a two level system (TLS) formed by an
electron in the ground state $|0 \rangle$ and a trion in the excited
state $|1 \rangle$. In the model developed below the coupling
between the optical fields and the TLS is treated exactly. The
optical response of the TLS is expressed at all orders of the
exciting field amplitudes (not limited to the third and five orders,
which are usually considered in the low field intensity regime). The
relevant semiclassical Hamiltonian of the  considered system in
interaction with the three exciting fields is given by $H =H_0+H_L,$

 \begin{gather}
  H_0= \hbar \omega_{eg}|1\rangle\langle 1|,\nonumber  \\
  H_L = \sum\limits_{i=1}^3 \mu \, \Ei (t)\,e^{i(\omega_{eg} (t-\tau_i)+\phi_i)}|0 \rangle\langle 1|+\text{c.c.},
 \end{gather}

where $\omega_{eg}$ refers to  the electron-trion optical transition
frequency, $\mu$ is the electric dipole of the QD, $\Ei (t)$
determines the effective electric field amplitude of the pulse $i$
at the  QD position, $\tau_i$ represents the arrival time of the
pulse $i$ at the QD position  and $\phi_i$ refers  to the additional
phase  shift induced by optical heterodyning. The excitation pulses
are assumed to be delta-function pulses, i.\,e., $\Ei (t)=\Ei
\delta(t-\tau_i)$. In addition to the evolution governed by the
Hamiltonian $H$, the system undergoes dissipative dynamics due to
the coupling  with its environment. Therefore the state of the
system $\rho(t) $ is described in the density matrix formalism in
the basis $\left\lbrace \Ket{0},\Ket{1}\right\rbrace $.

In order to relate the different multi-wave responses we derive the
state of the system $\rho(t) $. We assume that the pulse duration is
much shorter than the relevant time scales of the system dynamics.
Therefore the interaction between the system and its dissipative
environment is neglected during the coupling with the optical
excitation fields. Under this assumption, the modification of the
state $\rho(t)$ induced by the pulses is obtained
analytically\,$^{36,\,37}$ according to the unitary transformation
$U(\theta)$\,:

\begin{gather}
  \rho(\tau_{i}^+)=U(\theta_i) \rho(\tau_{i}^{-}) U(\theta_i)^{\dag}, \\
U(\theta_i)=\cos{\theta_i \over{2}}[|0\rangle\langle
0|+|1\rangle\langle 1|]-i\sin{\theta_i \over{2}}[|0 \rangle\langle
1|e^{i(\phi_i-\omega_{eg}\tau_i)} + c.c.],
 \end{gather}

where $\tau_i^{-(+)}$ denotes the time instant just before (after)
the pulse $i$, and   $\theta_i=\int\nolimits_{-\infty}^{+\infty} dt
\mu  \Ei (t)/ \hbar$ is the area of the pulse $i$. During the delay
time between   the pulses the open  two-level system relaxes its
density and polarization.

In the following the dissipation  is described in the Markov limit,
the subsequent dynamics can be then given  by  a master equation in
the Lindblad form
\begin{equation}
\dot{\rho}= - \frac{i}{\hbar} [H_0,\rho ]+\mathcal{L}(\rho)
\end{equation}
where $\mathcal{L}$ is the superoperator of dissipation. It takes
into account the dissipation of energy (mainly induced by
spontaneous emission in our experimental case) and pure dephasing
processes, which are characterized respectively by the rates
$\gamma_1$ and $\gamma_d$.  Given the minute values of fluctuations
induced by the spectral wandering (as measured in Fig.\,1.\,c), the
inhomogeneous dephasing is disregarded here. Therefore $\mathcal{L}$
reads as\,:

\begin{equation}
 \mathcal{L}(\rho)= \gamma_1[C\rho C^{\dag}-\frac{1}{2}\left\lbrace C^{\dag} C, \rho \right\rbrace_+ ]
  +\frac{1}{2} \gamma_d [D \rho D -\frac{1}{2} \left\lbrace D^2, \rho \right\rbrace_+  ]
\end{equation}

with $C= |1 \rangle\langle 0| \;  \text{and} \; D=|1 \rangle\langle
1|-|0 \rangle\langle 0|$.

By solving this equation we derive  the complete  optical
polarization  dynamics, which is proportional to the off-diagonal
terms of the density matrix. We  deduce  therefore the  field $\Ed$
radiated by the quantum dot structure\,:

\begin{equation}
 \Ed(t) = A\Bra{1} \rho(t) \Ket{0}+c.c. \,\,\,  \text{with}  \,\, A=i\mu_0 c \omega_{eg}\mu \kappa
\end{equation}

where $\kappa$  represents a coefficient of proportionality
including phase shift and absorption due to the propagation inside
the structure.

The coherent   optical response of  the  QD can be developed in a
sum of contributions $\Eg $, referring to particular nonlinear
processes. The emitted fields $\Eg $ are identified  by their phase
factors $e^{i\phi_{\boldsymbol{n}}}$, where the phase
$\phi_{\boldsymbol{n}}$ results  from a  linear combination of the
exciting pulse  phases  : $\phi_{\boldsymbol{n}}= \sum\nolimits_{i}
a_i \phi_i $ and $\boldsymbol{n}=\lbrace a_1,a_2,a_3\rbrace \in
\mathbb{Z}^3$.

The principle of the heterodyne spectral interferometry experiment
is based on the capacity to select a specific $\Eg$ via
corresponding phase $\phi_{\boldsymbol{n}}$. To do so, the phase
relation between the exciting pulses is tailored by acousto-optics
modulators operating at radio-frequencies $\Omega_i$. Therefore
successive exciting pulses $\lbrace\Ei\rbrace$ belonging to optical
train $i$, characterized by a repetition rate $\tau_r$, are phase
shifted by a factor $\Omega_i \tau_r$. By repeating the experiment
in time and combining optical heterodyning with the spectral
interferometry we retrieve the amplitude and phase of the desired
non-linear component $\Eg$ out from the total electric field
radiated by the QD dipole\,$^8$. In the present study, the nonlinear
signals of interest are the following\,:
\begin{itemize}
\item{ the non-degenerate FWM $\En$  beating at the heterodyne frequency $\phi_{\boldsymbol{f'}}=\phi_3- \phi_2 + \phi_1$}
\item{ the degenerate FWM  $\Ef$ beating at the heterodyne frequency  $\phi_{\boldsymbol{f}}=2\phi_2 - \phi_1 $}
\item{the degenerate SWM $\Es$ beating at the heterodyne frequency  $\phi_{\boldsymbol{s}}=2\phi_3- 2\phi_2 + \phi_1 $ }
\end{itemize}

Considering the TLS  in the ground state at the time $\tau_1^{-}$,
the dynamics of   the optical fields are given by\,:
 \begin{align}
\label{SM1}
& \En (t)=\frac{A}{2} \sin({\theta_1})\sin({\theta_2})\sin({\theta_3})  \mathrm{\Theta} (t-\tau_{23}) e^{-(t-\tau_{23}+\tau_{12})\gamma_2-\tau_{23} \gamma_1 }e^{i(\omega_{eg} (t-\tau_{23}+\tau_{12})+\phi_{\boldsymbol{f'}})}  \\
\label{SM2}
& \Ef (t)=A \sin({\theta_1})\sin^2({\frac{\theta_2}{2}}) \left(  \mathrm{\Theta} (t)-\sin^2({\frac{\theta_3}{2}}) \mathrm{\Theta} (t-\tau_{23}) \right) e^{-t \gamma_2 }e^{i(\omega_{eg} (t-\tau_{12})+\phi_{\boldsymbol{f}})}  \\
\label{SM3} & \Es (t) =-A \sin({\theta_1})
\sin^2({\frac{\theta_2}{2}}) \sin^2({\frac{\theta_3}{2}})
\mathrm{\Theta} (t-\tau_{23}) e^{-t \gamma_2 }e^{i(\omega_{eg}
(t-2\tau_{23}+\tau_{12})+\phi_{\boldsymbol{s}})},
\end{align}
where $\Theta(t)$ is the Heaviside function, $\gamma_2=\gamma_1 /
2+\gamma_d$ and  $(\tau_1,\,\tau_2,\,\tau_3)=(-\tau_{12},
0,\tau_{23}) $ with $\tau_{ij}=\tau_j-\tau_i$ and $\tau_{12}>0$ and
$\tau_{23}>0$. Once Fourier transformed with respect to the  time
\textit{t}, the spectrally resolved wave mixing field amplitudes
read as\,:
\begin{align}
\label{TF1}
& \En (\omega)=\frac{A}{2} \sin({\theta_1})\sin({\theta_2})\sin({\theta_3}) \, \, \frac {e^{-\tau_{12}\gamma_2-\tau_{23}\gamma_1 }}{\sqrt{2\pi}(\gamma_2-i\Delta \omega)} , \\
\label{TF2}
& \Ef (\omega)=A \sin({\theta_1})\sin^2({\frac{\theta_2}{2}}) \, \frac{ e^{-\tau_{12}{\gamma_2}} \left(1  -\sin^2({\frac{\theta_3}{2}})e^{\tau_{23}(i \Delta \omega -\gamma_2)}  \right)  }{\sqrt{2\pi}(\gamma_2-i\Delta \omega)},  \\
\label{TF3} & \Es (\omega) =-A \sin({\theta_1})
\sin^2({\frac{\theta_2}{2}}) \sin^2({\frac{\theta_3}{2}}) \, \,
\frac {e^{-(\tau_{12}+\tau_{23})\gamma_2
}}{\sqrt{2\pi}(\gamma_2-i\Delta \omega)} ,
\end{align}
where $\Delta \omega= \omega_{eg}-\omega$ and the phases
$\phi_{\boldsymbol{n}}$ are omitted. In the equations (\ref{SM1})
and (\ref{TF1}), one can note that the amplitude of the non
degenerate FWM signal $\En$ is sensitive to both the relaxation of
the polarization and density of the TLS. In particular, the
expression (\ref{TF1}) in the spectral domain  highlights that
probing the signal $\En$ as a function of the delay times
$\tau_{12}$ and $\tau_{23}$ allows to experimentally determine the
characteristic rates $\gamma_1$ and $\gamma_2$, respectively (see
Fig.\,1 in the main manuscript and Supplementary Fig.\,S4).

The relations (\ref{SM2}) and (\ref{SM3}) are describing the core of
the experiments presented in the main article. They explicitly
illustrate the dynamics of the TLS coherences induced by the third
pulse, where the FWM response is projected to the SWM response. This
interplay between the two nonlinear processes is controlled by the
area $\theta_3$ of the third pulse and the time delay $\tau_{23}$
between the last two pulses. Conversely, such degree of freedom
permits designing the spectro-temporal shape of the FWM component,
as shown in the equations (\ref{SM2},\,\ref{TF2}).

We point out that the switching between FWM and SWM triggered by
$\Ec$, physically modifies the coherent oscillation of the dipole
and directly acts on the total lineshape of its emission. This is
exemplified in Supplementary Figure
Fig.\,\ref{fig:SM_globalemission} (see page 9). Therein, we present
simulated spectral lineshape of the complete QD optical polarization
after the arrival of the second pulse $\Eb$ for
$(\theta_1,\,\theta_2,\,\theta_3)=(\frac{\pi}{2},\,\pi,\,\pi)$. We
observe that the spectral lineshape of the dipole is significantly
altered (for example for $\tau_{23}=390\,$ps) compared to the
unperturbed case ($\theta_3=0$). The spectral broadening is
generated by the spectral interferences between the FWM (for $t <
\tau_{23}$) and SWM (for $t> \tau_{23}$) components, which averaged
over many repetitions of the heterodyne experiment produce such
spectrum.

\section{Experimental details}
Our experiments are based on an extended version of the heterodyne
spectral interferometry technique\,$^8$. We use a pulse train
spectrally centered at $\hbar\omega_0=1.35eV$, generated by a
Ti:Saphire laser (Spectra-Physics, Tsunami Femto). The pulses are
spectrally shaped and chirp-corrected by a passive pulse-shaper (G -
grating, LS - lens, SM - spherical mirror). They are then spectrally
up-shifted by three acousto-optic modulators (AOMs) operating at
$\Omega_{1,2,3}=(80,\,79,\,80.77)\,$MHz. Thus created three driving
fields $\Eo$ - marked as red, blue and green, respectively - acquire
relative time delay $\tau_{13}$ (positive for $\Ea$ leading) and
$\tau_{12}$ (positive for $\Ea$ leading) by passing through two
mechanical delay lines. Their polarization can be controlled by a
set of half- and quarter-plates, but in this work we employ
co-linear, horizontal (in the lab-frame) polarization for all beams.
The driving fields are recombined into the same spatial mode
(yellow) and focussed on the sample surface via the microscope
objective (MO - Olympus, LCPLN50XIR/0.65, BS –beam sampler, FF – far
field) installed on a X-Y-Z close-loop piezo-stage, (XY-range of
250\,$\mu$m, Z- range of 450\,$\mu$m). The excitation pulses have
duration of $350\pm50$\,fs. The pulse chirp, mainly induced by a
slab of TeO$_2$ crystal in the AOMs and in the MO, is compensated by
measuring the non-resonant four-wave mixing produced at pulse
overlap in an auxiliary GaAs sample and minimizing its duration with
respect to $\tau_{12}$. The microcavity sample is installed inside a
custom, continuous flow helium cryostat, containing motorized,
closed-loop XY translation stages (range of 10\,mm, accuracy of
50\,nm). Experiments reported here were performed at
$5.1\,\pm0.1\,$K. The far field is imaged into the AOM, where -
using a Bragg condition - the reflected signal is mixed with the
reference field $\Er$. The mixing AOM is driven at the frequency
$\Omega_{\rm{D}}= a\,\Omega_1+ b\,\Omega_2+ c\,\Omega_3$, where
$(a,\,b,\,c)\,\epsilon\,(-3,\,-2,\,-1,\,0,\,1,\,2,\,3)$ and a+b+c=1.
These heterodyne frequencies are produced by the home-made
three-channel, analogue mixer, with a spectral purity of 40\,dB.
They are subsequently regenerated by a lock-in amplifier attaining
spectral purity of 100\,dB. Mixed beams $\rm{W}_{\rm{A}}$ and
$\rm{W}_{\rm{B}}$ are imaged into the entrance slit of the imaging
spectrometer of 750\,mm focal length. Spectral interferograms are
retrieved by subtracting and $\pi$-phase flipping $\rm{W}_{\rm{A}}$
and $\rm{W}_{\rm{B}}$, as described in Ref.\,[8]. We use a CCD
camera from Princeton Instruments (PIXIS:400BR eXcelon), offering
enhanced full well capacity, readout rate and quantum efficiency of
83\% at 915nm. The wave mixing signal is retrieved in amplitude and
phase by performing spectral interferometry and by adjusting $\Er$
to arrive prior to the signal. We note that we have also tested
optical heterodyning by performing digital mixing with an arbitrary
wave-form generator (Tektronix, sampling rate 1.2G/s)  and retrieved
the FWM from single quantum dots with the same signal-to-noise
ratio, as with the analogue mixing. This paves the way towards
heterodyne n-wave mixing experiments, enabling quantum control of
high order nonlinear processes of individual optical transitions in
solids.

\begin{figure}[h!]
\includegraphics*[width=1.02\columnwidth]{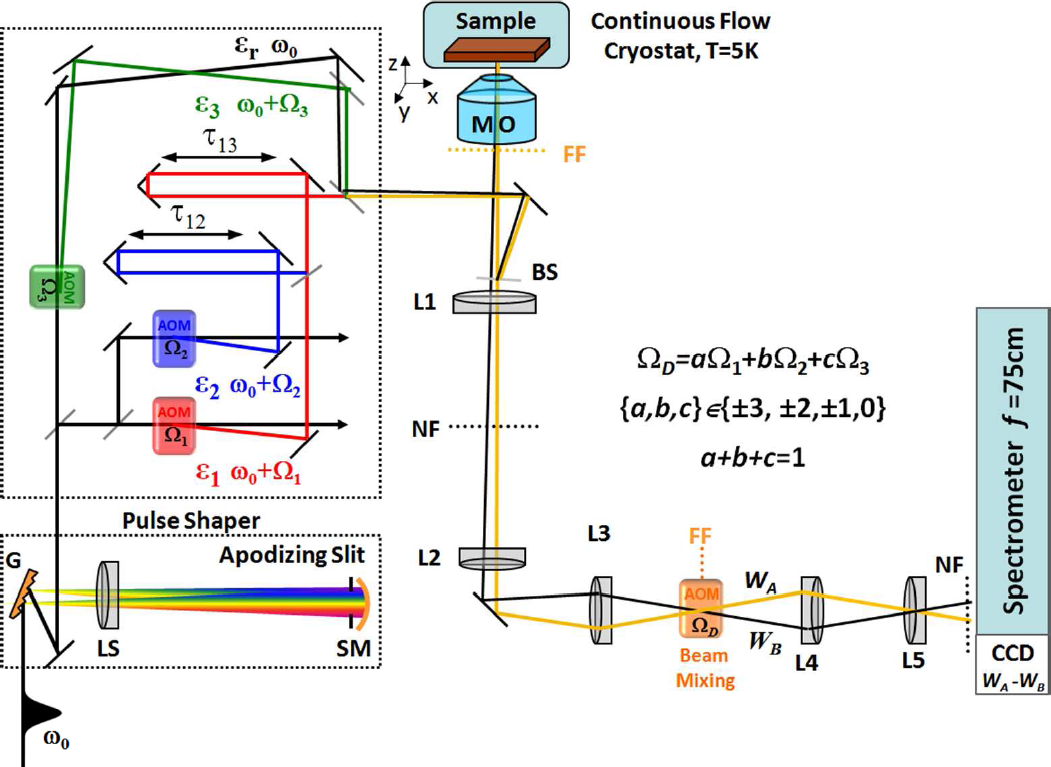}
\caption{{\bf Schema of the experimental setup.}
\label{fig:SM_setup}}
\end{figure}

\section{Auxiliary results}

\begin{figure}[h!]
\includegraphics*[width=1.02\columnwidth]{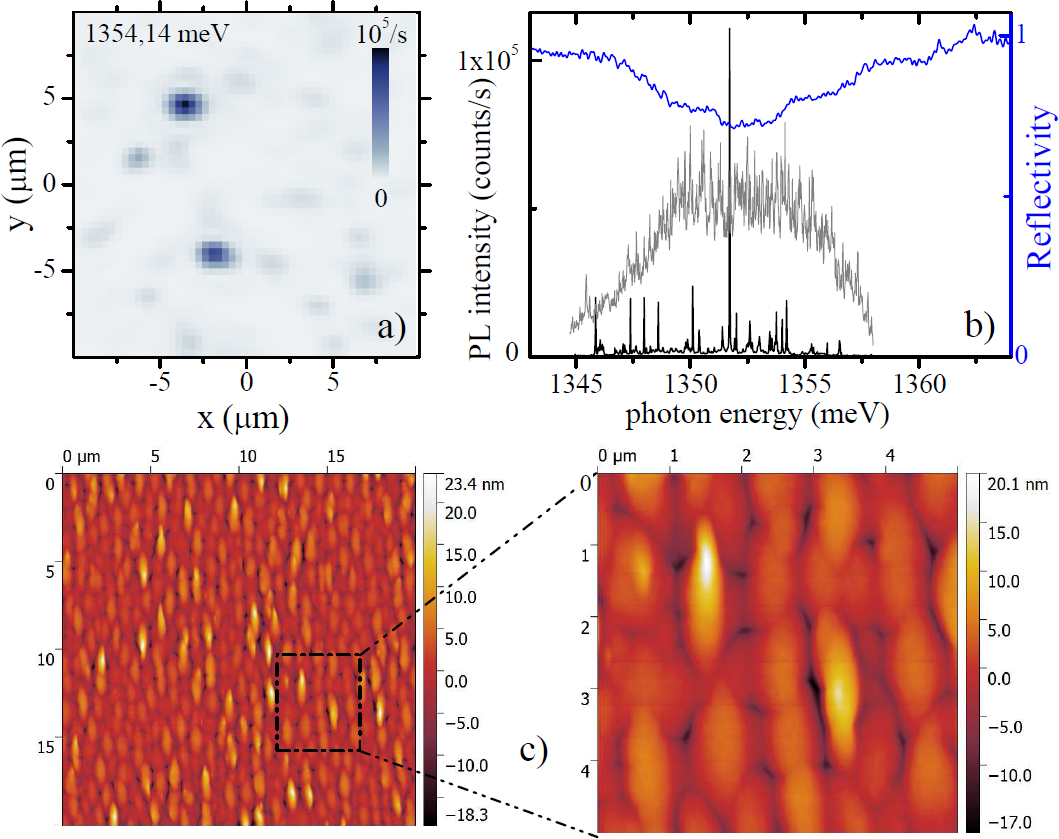}
\caption{{\bf Sample characterization: micro-photoluminescence,
reflectivity and surface topography.} \label{fig:SM_SM_PL_Refl_AFM}}
\end{figure}

The sample has been grown by the molecular beam epitaxy. A layer of
annealed and capped InAs QDs (density $2\times10^9\,$cm$^{-2}$) is
placed in a center of a GaAs spacer. A $\delta$-doping layer with Si
(density $1.8\times10^{10}\,$cm$^{-2}$) is present 10\,nm below the
QD layer, inducing intentional negative doping of QDs. The spacer is
sandwiched between two Bragg mirrors. The bottom (top) mirror
contains 24 (5) pairs of $\lambda/4$ Al$_{0.9}$GaAs/GaAs stacks,
forming an asymmetric $\lambda$-cavity with a quality factor of
$Q\simeq170$. An example of the micro-photoluminescence
hyperspectral imaging frame performed on the studied microcavity is
shown in Fig.\,\ref{fig:SM_SM_PL_Refl_AFM}\,a. The sample is excited
in continuous wave at 1.734\,eV with $0.3\,\mu$W at the sample
surface. Blue peaks correspond to the emission of individual InAs
quantum dots. Count rates at the peak as high as $10^5$ per seconds
are routinely detected using a grating of 1800 grooves/mm with Low
Noise and gain 2 settings of the Pixis camera. Linear color scale
from 0 to $10^5$ counts per seconds. In
Fig.\,\ref{fig:SM_SM_PL_Refl_AFM}\,b we present a typical
micro-photoluminescence spectrum (black) and photoluminescence
spectrum averaged over $400\,\mu\rm{m}^2$ (gray). The reflectivity
spectrum measured with the white light (blue) shows the cavity mode.
The mode is broadened at the high energy side due to high NA of the
microscope objective and the in-plane dispersion of the cavity. In
Fig.\,\ref{fig:SM_SM_PL_Refl_AFM}\,c we present topography of the
sample surface, revealed with the atomic force microscopy. Typical
topography images over the sample area of $20\,\mu\rm{m}\,\times\,20
\,\mu\rm{m}$ (left) and of $5\,\mu\rm{m}\,\times\,5 \,\mu\rm{m}$ are
shown, revealing mosaic-like morphology with characteristic oval
photonic defects of typically $2\,\mu\rm{m}\,\times\,1 \,\mu\rm{m}$
in-plane size and of 20\,nm hight occurring with a spatial density
0.1$\mu \rm{m}^{-1}$. Such defects offer an enhanced in- and
out-coupling efficiency of around 40\,\%, as recently shown in
Ref.\,[20]. This, combined with an intra-cavity amplification,
results in an enhanced excitation and detection of the coherent
nonlinear responses of QDs situated close to the maximum of the
localized optical mode induced by a defect. We systematically
measure shorter T$_1$ times (150-350\,ps), weaker driving powers and
stronger FWM from QDs residing in the defects, with respect to the
dots placed outside of defects. We speculate that these observations
are due to the change of the photon density of states by the surface
modulation defects, creating a varying coupling efficiency and
Purcell effect. Further research is required to make a clear
assignment and to elucidate the influence of these photonics defects
on the coherent response of QDs.

\begin{figure}[h!]
\includegraphics*[width=0.85\columnwidth]{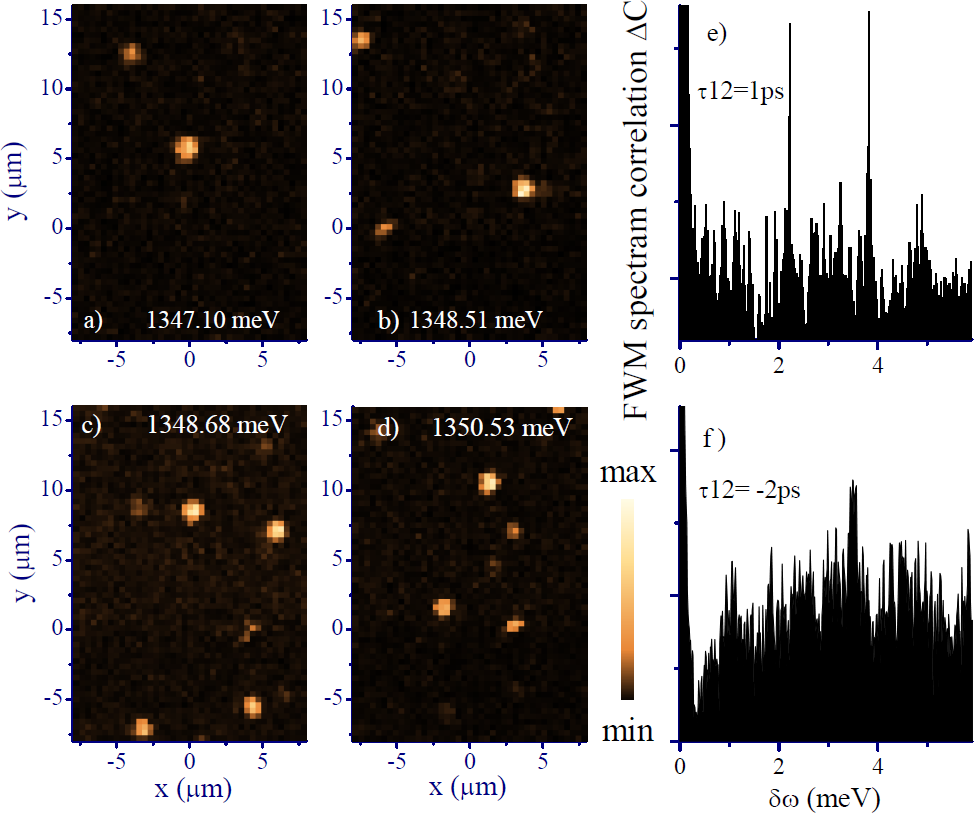}
\caption{{\bf FWM hyperspectral imaging and correlation analysis.}
 \label{fig:SM_correlation} }
\end{figure}

Typical frames of the FWM hyperspectral imaging at
$\tau_{12}=+1\,\rm{ps}$ and $\omega$ as labeled are shown in
Fig.\,\ref{fig:SM_correlation}\,a-d. Linear colour scale from black
to white, as shown by the vertical bar. Strongest peaks primarily
correspond the FWM of negative trions. In order to obtain
spectrally-resolved FWM image, as exemplified in
Fig.\,\ref{fig:WM_spec}\,b, the following procedure is applied. For
each spatial position, the spectrally-resolved FWM amplitudes are
fitted with a Lorentzian function within a chosen energy region
(specifically, from 1349.7\,meV to 1350.3\,meV in
Fig.\,\ref{fig:WM_spec}\,b). The fitted center position yields the
energy, while the fitted Lorentzian amplitude is proportional to the
FWM amplitude. The retrieved parameters are rearranged into a pair
of two-dimensional spatial maps; encoding spectral position and
amplitude, respectively. They are finally merged into a single,
spatial map, where the amplitude is represented as brightness and
the energy as a hue level.

In Fig.\,\ref{fig:SM_correlation}\,e and f we provide the result of
the FWM spectral correlation analysis $\Delta C(\delta\omega)$
performed on the hyperspectral imaging FWM data, as defined in
Ref.\,[38]. For $\tau_{12}=+1\,$ps, $\Delta C(\delta\omega)$ reveals
two peaks at $\delta\omega$ of 2.2\,meV and 3.8\,meV attributed to
neutral exciton and biexciton transitions, present in around 10\% of
quantum dots. Due to the weak strength of two particle states,
specifically negatively charged biexcitons, the FWM of trions is
vanishing at negative delays $\tau_{12}<0$. Therefore for
$\tau_{12}=-2\,$ps, $\Delta C(\delta\omega)$ shows only one peak at
$\delta\omega=3.6\,$meV, attributed to biexciton transitions of
neutral excitons. Lower signal to noise ratio is due to smaller
statistics available. Wave mixing spectroscopy on such
exciton-biexciton-trion systems will be presented in a forthcoming
publication.

\begin{figure}[h!]
\includegraphics*[width=0.85\columnwidth]{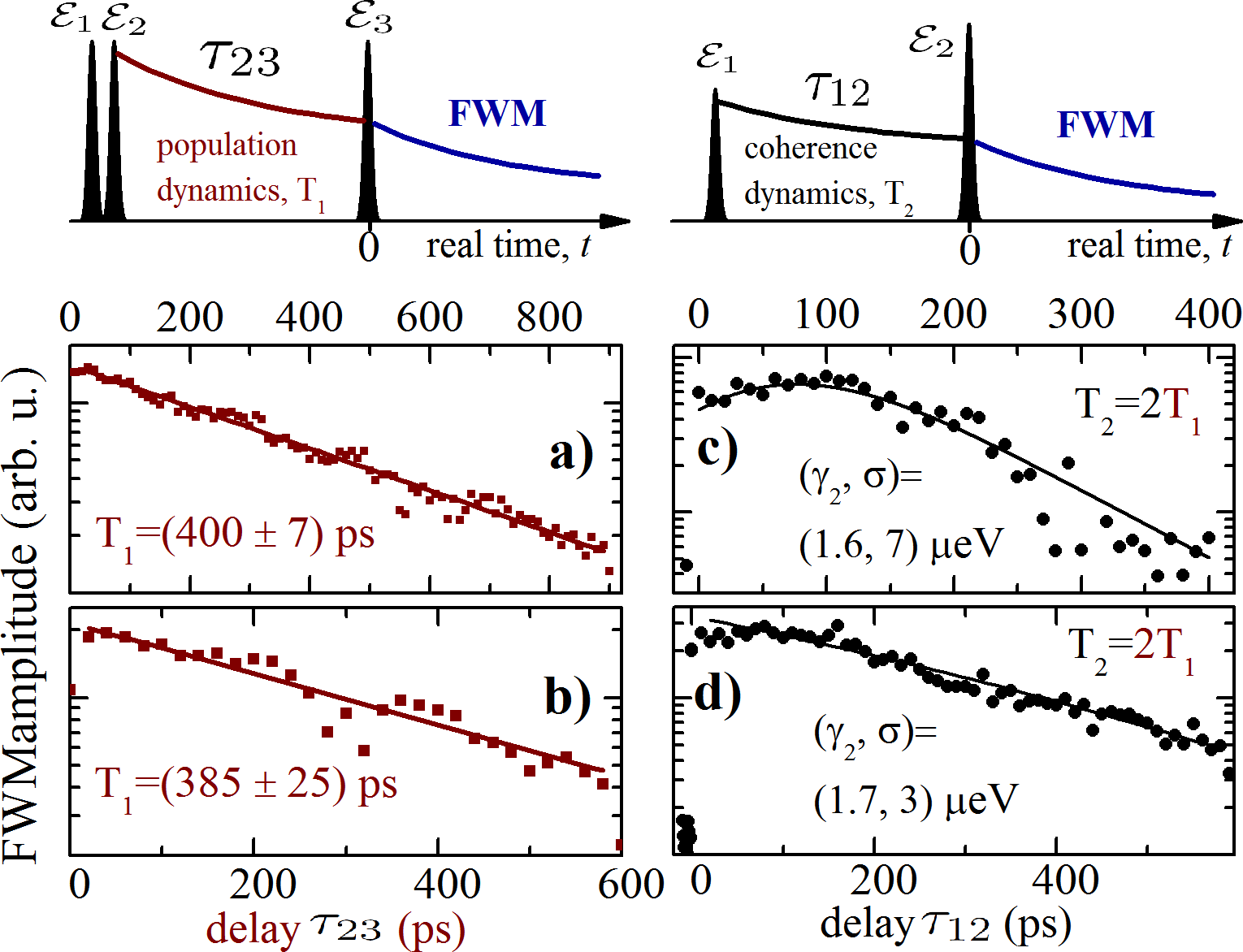}
\caption{{\bf Radiatively limited dephasing measured on individual
negative trions. Applied pulse sequences are depicted at the top.}
\label{fig:SM_dephasingtrions}}
\end{figure}

In Fig.\,\ref{fig:SM_dephasingtrions} we provide two supplementary
FWM measurements performed on two different negative trions,
demonstrating their raditively limited dephasing (T$_2$=2T$_1$). The
population dynamics measured with FWM is shown by brown squares.
Fitted mono-exponential decay (brown line) yields population
lifetime of T$_1$=$(400\,\pm\,7)\,$ps (a) and
T$_1$=$(385\,\pm\,25)\,$ps (b). The coherence dynamics measured by
FWM is depicted with black circles in (c) and (d), respectively. The
data are fitted by the model presented in Ref.\,[19] (namely
Eq.\,(2)), using radiatively limited dephasing and inhomogeneous
broadening due to residual spectral wandering $\sigma=7\,\mu$eV (c)
and $\sigma=3\,\mu$eV (d). Note the initial rise in the coherence
dynamics observed in (c), due to the resolved photon echo formation.
This figure is supplementary to Fig.\,1. Pulse sequences employed to
measure coherence (right) and population (left) dynamics are also
depicted. The reference pulse $\Er$ arrives a few ps prior to $t=0$.

\begin{figure}[h!]
\includegraphics*[width=0.8\columnwidth]{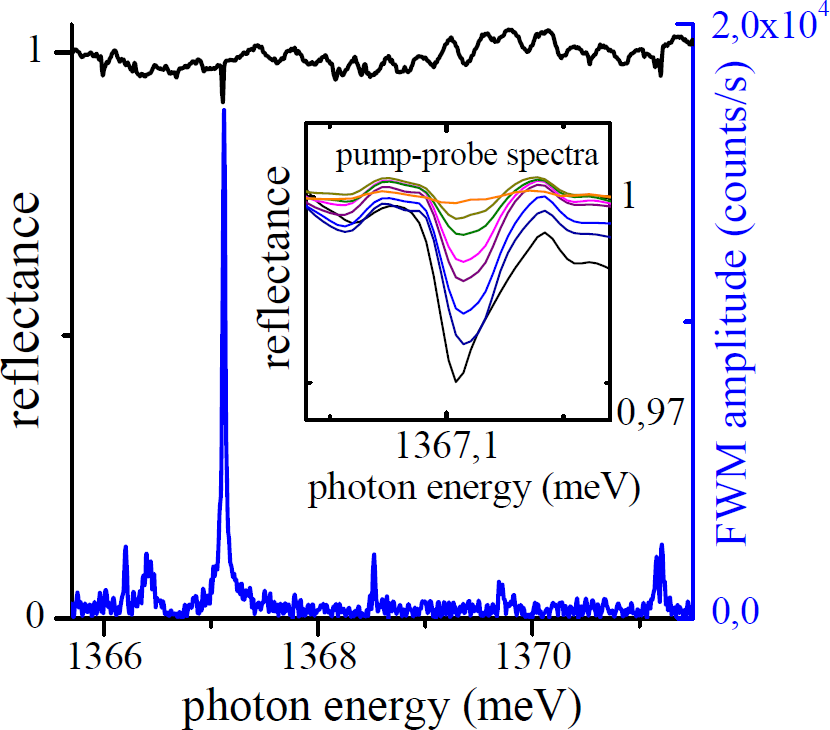}
\caption{{\bf Signatures of single trions observed in the direct
reflectance and the corresponding pump-probe spectra measured in
reflectance.} \label{fig:SM_refl}}
\end{figure}

Being radiatively limited, such trions are expected to be observable
in the first-order absorption experiment. In Fig.\,\ref{fig:SM_refl}
we compare the FWM spectrum (blue line) with a direct reflectance
spectrum of the $\Ec$ beam (black line) measured at the same spot.
The FWM reveals dominating transition at 1367.1\,meV. The
reflectance displays a dip at the same spectral position. It
corresponds to the first-order absorption of the QD, with a contrast
of a few per-cent resulting from the convolution of the homogenous
linewidth with the spectral resolution. To further confirm that the
reflectance dip indeed corresponds to the linear absorption of a
trion in a single QD, we have performed the pump-probe experiment.
The result is given in the inset, where the $\Eb$ reflectance is
shown while increasing the pulse area of the $\Ea$ beam from
$\theta_1=0$ (black trace) to $\theta_1=\pi$ (orange trace) for
$\tau_{12}=2\,$ps ($\Ea$ arriving first). As expected, the $\Eb$
absorption is gradually decreased with increasing $\theta_1$ and is
entirely saturated for the $\theta_1=\pi$.

\begin{figure}[h!]
\includegraphics*[width=0.7\columnwidth]{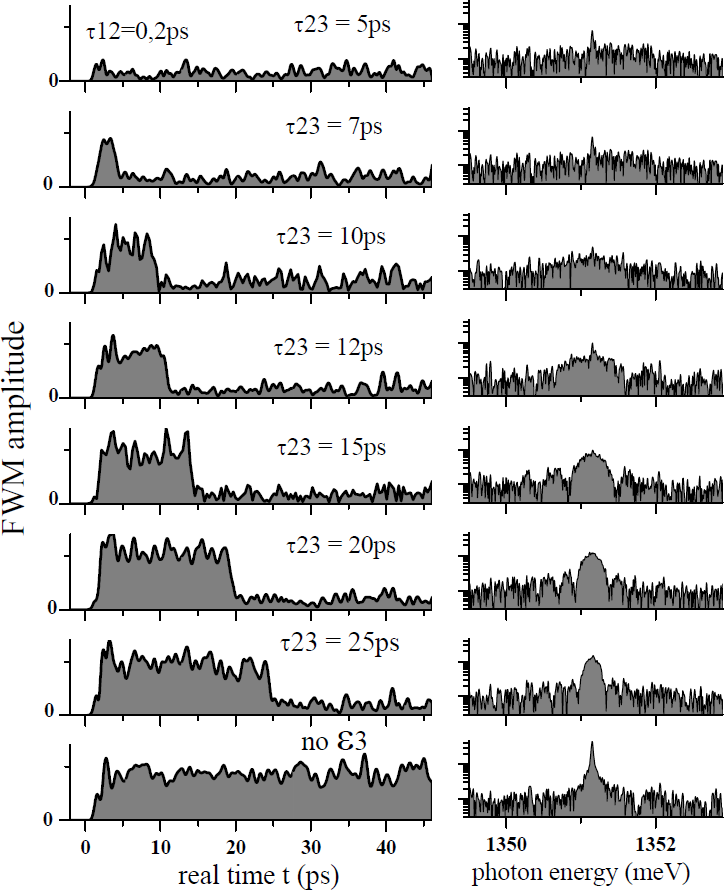}
\caption{{\bf Control of the FWM transient with the delay
$\tau_{23}$, between $\Eb$ of $\Ec$.} \label{fig:SM_controlFWM}}
\end{figure}

\begin{figure}
\includegraphics*[width=0.8\columnwidth]{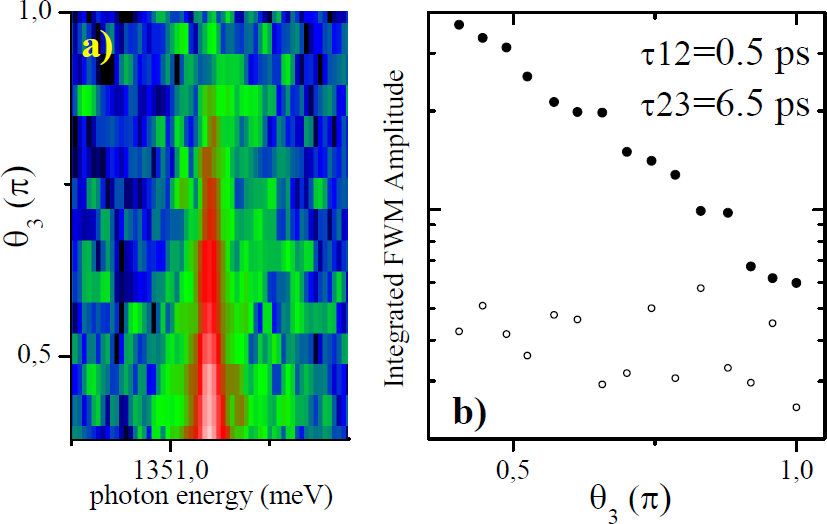}
\caption{{\bf Control of the FWM with the pulse area of $\Ec$.}
 \label{fig:SM_controlFWM_pulsearea}}
\end{figure}

In Fig.\,\ref{fig:SM_controlFWM} we present the amplitude of the FWM
transient for $\tau_{12}=0.2\,\rm{ps}$, demonstrating a complete
suppression of the FWM at the arrival of the $\Ec$, which converts
the FWM polarization into the SWM one. The pulse area of $\Ec$ is
set to $\pi$. Right: Measured FWM spectrum, showing the
corresponding change of the FWM spectral lineshape depending on
$\tau_{23}$. The data are supplementary to Fig.\,3. In
Fig.\,\ref{fig:SM_controlFWM_pulsearea}, the FWM is generated only
within a time window of $\tau_{23}=6.5\,$ps prior to the arrival of
$\Ec$. As a result, with increasing $\theta_3$, the FWM gets
virtually suppressed. The data are supplementary to Fig.\,4.

\begin{figure}
\includegraphics*[width=0.7\columnwidth]{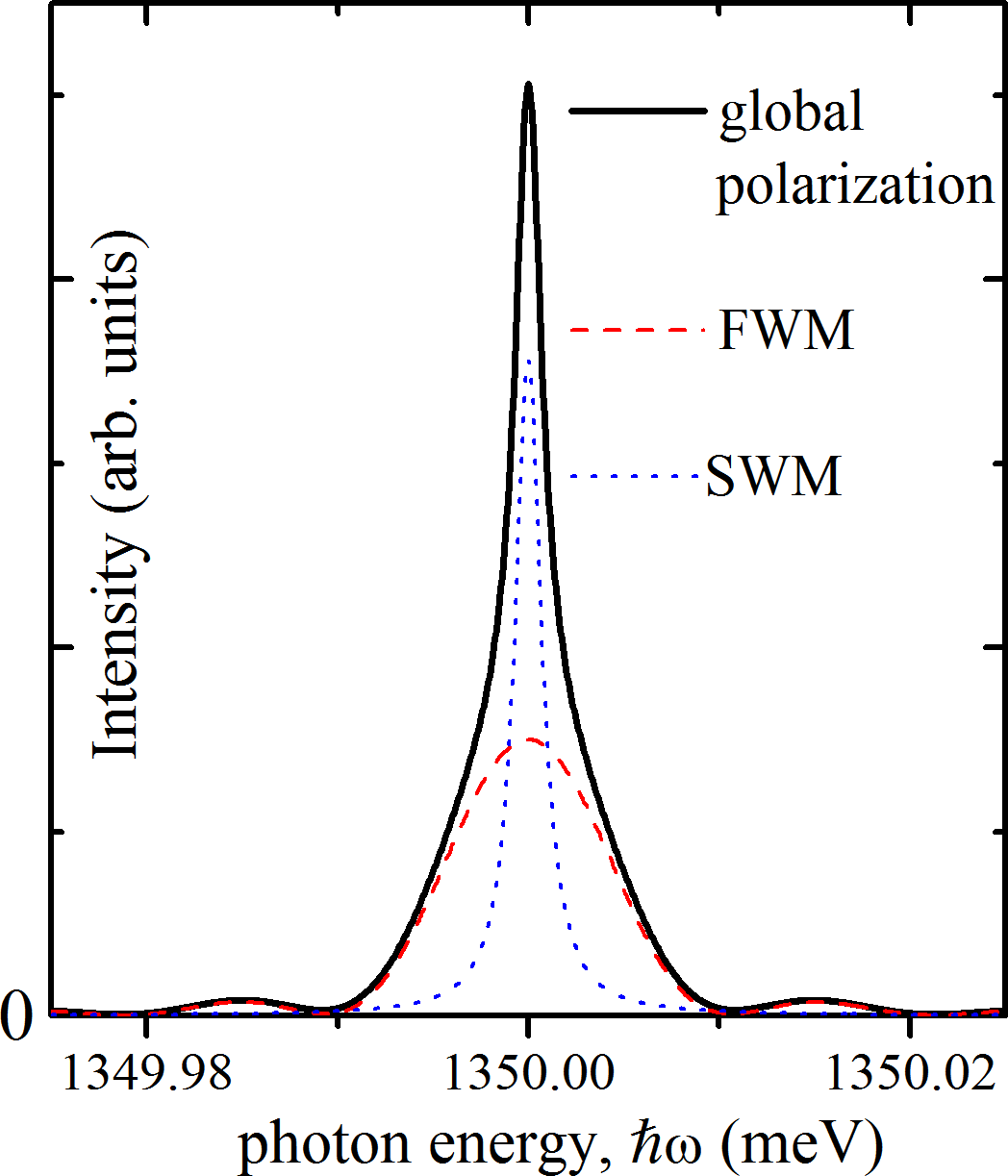}
\caption{{\bf Modification of the global lineshape of a TLS via
FWM/SWM switching.} Spectrum of the complete dipole polarization
(black solid line) after the arrival of the second pulse $\Eb$ for
$(\theta_1,\,\theta_2,\,\theta_3)=(\frac{\pi}{2},\,\pi,\,\pi)$,
$T_2=780$\,ps and $\tau_{23}=390\,$ps. For such value of
$\tau_{23}$, the integrated amplitudes of the FWM and SWM components
are almost equivalent, which maximizes the spectral interferences
between them. FWM (red dashed line) and SWM (blue dotted line)
spectra are presented as well. \label{fig:SM_globalemission}}
\end{figure}

\section*{Supplementary References}

$^{36}$ A. Bambini and P. R. Berman, Analytic solutions to the
two-state problem for a class of coupling potentials \emph{Phys.
Rev. A} {\bf 23}, $2496$ (1981).

$^{37}$ Lorenza Viola and Seth Lloyd, Dynamical suppression of
decoherence in two-state quantum systems \emph{Phys. Rev. A} {\bf
58}, $2733$ (1998).

$^{38}$ J.\,Kasprzak and W.\,Langbein, Coherent response of
individual weakly confined exciton–biexciton systems \emph{J. Opt.
Soc. Am. B} {\bf 29}, 1776 (2012).

\end{document}